\documentclass[twocolumn,prb,showpacs,superscriptaddress,preprintnumbers,amsmath,amssymb,floatfix]{revtex4}
\usepackage{graphicx}
\usepackage{dcolumn}
\usepackage{color}

\usepackage{bm}

\begin{document}

\title{Intrinsic and extrinsic x-ray absorption effects in soft x-ray diffraction from the superstructure in magnetite}

\author{C.~F.~Chang}
  \affiliation{II. Physikalisches Institut, Universit\"{a}t zu K\"{o}ln, Z\"{u}lpicher Str. 77, 50937 K\"{o}ln, Germany}
\author{J.~Schlappa}
  \affiliation{II. Physikalisches Institut, Universit\"{a}t zu K\"{o}ln, Z\"{u}lpicher Str. 77, 50937 K\"{o}ln, Germany}
\author{M.~Buchholz}
  \affiliation{II. Physikalisches Institut, Universit\"{a}t zu K\"{o}ln, Z\"{u}lpicher Str. 77, 50937 K\"{o}ln, Germany}
\author{A.~Tanaka}
  \affiliation{Department of Quantum Matter, ADSM, Hiroshima University, Higashi-Hiroshima 739-8530, Japan}
\author{E.~Schierle}
  \affiliation{Helmholtz-Zentrum Berlin, Albert-Einstein-Str. 15, 12489 Berlin, Germany}
\author{D.~Schmitz}
  \affiliation{Helmholtz-Zentrum Berlin, Albert-Einstein-Str. 15, 12489 Berlin, Germany}
\author{H.~Ott}
  \affiliation{II. Physikalisches Institut, Universit\"{a}t zu K\"{o}ln, Z\"{u}lpicher Str. 77, 50937 K\"{o}ln, Germany}
\author{R.~Sutarto}
  \affiliation{II. Physikalisches Institut, Universit\"{a}t zu K\"{o}ln, Z\"{u}lpicher Str. 77, 50937 K\"{o}ln, Germany}
\author{T.~Willers}
  \affiliation{II. Physikalisches Institut, Universit\"{a}t zu K\"{o}ln, Z\"{u}lpicher Str. 77, 50937 K\"{o}ln, Germany}
\author{P.~Metcalf}
  \affiliation{Department of Chemistry, Purdue University, West Lafayette, Indiana 47907, USA}
\author{L.~H.~Tjeng}
  \affiliation{II. Physikalisches Institut, Universit\"{a}t zu K\"{o}ln, Z\"{u}lpicher Str. 77, 50937 K\"{o}ln, Germany}
  \affiliation{Max Planck Institute for Chemical Physics of Solids, N\"othnitzerstr. 40, 01187 Dresden, Germany}
\author{C.~Sch{\"u}{\ss}ler-Langeheine}
  \affiliation{II. Physikalisches Institut, Universit\"{a}t zu K\"{o}ln, Z\"{u}lpicher Str. 77, 50937 K\"{o}ln, Germany}

\date{\today}

\begin{abstract}
We studied the (00$\frac{1}{2}$) diffraction peak in the
low-temperature phase of magnetite (Fe$_3$O$_4$) using resonant
soft x-ray diffraction (RSXD) at the Fe-$L_{2,3}$ and O-$K$
resonance. We studied both molecular-beam-epitaxy (MBE) grown
thin films and \textit{in-situ} cleaved single crystals. From the comparison we have been able to determine quantitatively the contribution of intrinsic absorption effects, thereby arriving at a consistent result for the (00$\frac{1}{2}$) diffraction peak spectrum. Our data also allow for the identification of extrinsic effects, e.g. for a detailed modeling of the spectra in case a "dead" surface layer is present that is only absorbing photons but does not contribute to the scattering signal.
\end{abstract}


\maketitle Magnetite, Fe$_{3}$O$_{4}$, shows a first order anomaly
in the temperature dependence of the electrical conductivity at
$T_{V}\sim$ 120 K, the famous Verwey transition.\cite{Verwey1939}
It is accompanied by a structural phase transition from the cubic
inverse spinel to a distorted structure. One usually connects this
transition with charge ordering of the Fe$^{2+}$ and Fe$^{3+}$
ions on the octahedrally coordinated, so-called B-sites. In a
recent diffraction study, Wright, Attfield and Radaelli found
long-range charge order from the pattern of shorter and longer
bond lengths between B-site iron and oxygen ions below $T_{V}$.
\cite{PhysRevLett.87.266401,PhysRevB.66.214422} Subsequently,
using this structure as input, local density approximation +
Hubbard $U$ (LDA+$U$) band structure studies have calculated the
corresponding orbital ordering, which involves mainly the minority
spin electron in the $t_{2g}$ state of the 2+ B-sites.
\cite{PhysRevLett.93.146404,PhysRevLett.93.156403}

Recently this $t_{2g}$ orbital order has been studied using RSXD
at the O $1s \rightarrow 2p$ ($K$) \cite{huang:096401} and Fe $2p
\rightarrow 3d$ ($L_{2,3}$) \cite{schlappa:026406} resonance. The
O-$K$ edge resonance enhancement of the (00$\frac{1}{2}$)
diffraction peak (notation refers to the cubic room-temperature
unit cell) on an \textit{ex-situ} polished bulk single crystal was
interpreted by Huang \textit{et al.} as a signature of a
particular charge/orbital order at the
oxygen-sites.\cite{huang:096401} At the Fe $L_{2,3}$-edges, using
MBE-grown thin films, the (00$\frac{1}{2}$) maximum energy
coincides with the resonance of the 2+ B-site ions and was
assigned by Schlappa \textit{et al.} to Fe $t_{2g}$-orbital order
on these sites.\cite{schlappa:026406}

Very recently these results have been challenged. Wilkins and coauthors have presented RSXD data from the (00$\frac{1}{2}$) peak [(001) in the orthorhombic notation used in Ref.~\onlinecite{wilkins:201102}] of bulk magnetite at the O-$K$ and Fe-$L_{2,3}$ resonance.\cite{wilkins:201102}. Their O-$K$ spectrum is similar to that of Huang \textit{et al.}\cite{huang:096401} but their Fe-$L_{2,3}$ data are
very different from those of Schlappa \textit{et al.}\cite{schlappa:026406} Almost at the same time Garc\'{i}a \textit{et al.}\cite{garcia:PRL2009} reported a Fe-$L_{2,3}$ RSXD spectrum from a bulk sample, which is very different from those of both Wilkins \textit{et al.} and Schlappa \textit{et al.}. In both recent publications\cite{wilkins:201102,garcia:PRL2009} the shape of the respective Fe-$L_{2,3}$ spectra are explained by absorption effects. Presently, hence, at least three different Fe-$L_{2,3}$ RSXD spectra for the (00$\frac{1}{2}$) peak of magnetite have been published and very different conclusions have been drawn about the electronic origin of the peak. 

Here we report our efforts to resolve the confusion about the RXSD
data on magnetite. Our strategy is to carry out the experiments
using a magnetite bulk single crystal which was carefully cleaved
\textit{in-situ} in order to obtain the best possible surface quality, and to compare the results with the measurements on high quality MBE-grown magnetite thin films. In this manner we are able to distinguish intrinsic absorption effects from extrinsic effects. Taking intrinsic effects properly into account, we find consistent results for \textit{in-situ} cleaved bulk samples and thin films. Furthermore, taking a so-called "dead" surface into account, which leads to extrinsic absorption, we are able to provide a quantitative explanation for the varying RSXD spectra reported in the literature so far.

The soft x-ray scattering experiments were performed at the HZB
beam line UE46-1 at BESSY using the two-circle UHV diffractometer
designed at the Freie Universit\"{a}t Berlin. We used 40 nm
thick magnetite films epitaxially grown on an epi-polished MgO(001)
substrates and several magnetite bulk single crystals. The samples
were oriented such that two base vectors of the cubic
room-temperature unit cell ($a=8.396$ \AA) were parallel to the
scattering plane. The polarization of incoming light was either
parallel or perpendicular to the scattering plane ($\pi$- or
$\sigma$-polarization, respectively). A silicon-diode photon
detector without polarization analysis was used with the angular
acceptance set to 1$^{\circ}$ in the scattering plane and
5$^{\circ}$ perpendicular to it. The energy dependence of the
diffraction-peak intensities (diffraction spectra) from the film
sample were recorded by varying the photon energy and keeping the
momentum transfer constant. For bulk samples this method leads to
distorted spectra because of the much narrower peaks and
refraction effects. We therefore performed longitudinal ($L$)
scans at every photon energy and integrated the
(00$\frac{1}{2}$)-peak intensity.

\begin{figure}
    \centering
    \includegraphics[width=6.7cm]{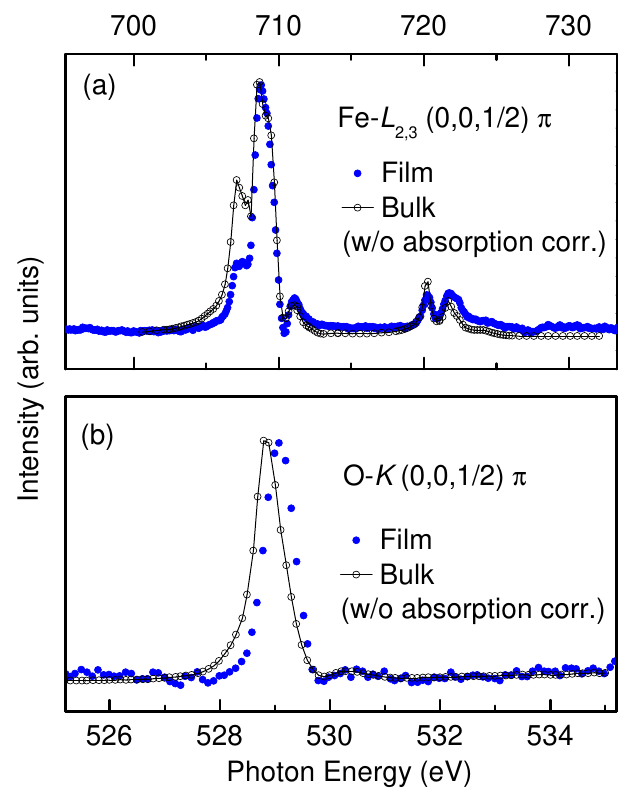}
    \caption{(color online) RSXD spectra of
    the (00$\frac{1}{2}$) peak of bulk magnetite (open symbols) and magnetite
    thin film (filled symbols) at (a) Fe $L_{2,3}$-edges and at (b) O
    $K$-edge. The spectra are scaled to equal height.}
    \label{Fig-bulk-film-NoCorr}
\end{figure}

Fig.~\ref{Fig-bulk-film-NoCorr} displays the (00$\frac{1}{2}$)
resonance spectra at the Fe $L_{2,3}$ and at O $K$-edges for bulk
magnetite and magnetite thin film. One can observe very distinct
differences between the bulk and the thin film. At the Fe edge,
for example, the first peak at 707 eV is much more pronounced in
the bulk than in the thin film. At the O edge, the energy position
of the peak of the bulk is clearly shifted with respect to that of
the thin film. We ascribe these differences to a change of the
scattering volume due to variations of the penetration depth of
the light as a function of its energy. This effect is quite severe
for energy variations across resonances in the soft
x-ray regime due to the very high photo-absorption coefficients.
For a bulk sample, the scattered intensity in specular geometry is proportional to
$1/\mu$, with $\mu$ being the absorption coefficient.
\cite{Als-Nielsen-McMorrow} This absorption effect
can be accounted for by multiplying the resonance spectra by
$\mu$; $\mu$ can be reliably obtained from the x-ray
absorption spectroscopy (XAS) signal recorded in the total
electron yield mode.

\begin{figure}[t]
    \centering
    \includegraphics[width=6.7cm]{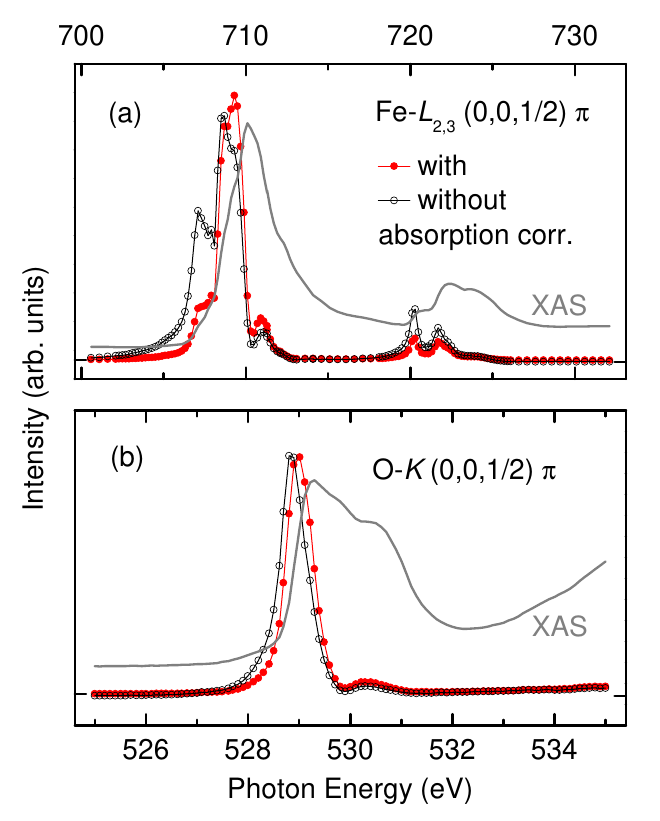}
    \caption{(color online) RSXD spectra of
    the (00$\frac{1}{2}$) peak of bulk magnetite without (open symbols) and with
    absorption correction (filled symbols) at (a) Fe $L_{2,3}$-edges and at (b) O
    $K$-edge. The spectra are scaled to equal height.
    XAS spectra are shown as solid lines.}
    \label{Fig-bulk-xasCorr}
\end{figure}

The effect of this correction is shown for bulk magnetite in
Fig.~\ref{Fig-bulk-xasCorr} for both the Fe-$L_{2,3}$ and O-$K$
resonances. The open symbols are the raw data and the filled
symbols are those after correction. The absorption signals are shown as solid lines. The correction for the Fe-spectrum
affects mainly the relative intensities of the spectral features,
while in the O-spectrum, the main peak, which is just on the slope
of the absorption signal, becomes shifted in energy. 

For a film of thickness $D$ the factor to correct for absorption effects is $\mu
/ (1-\exp(-2 \mu d))$. $d=D/\sin \theta$ describes the effective photon path length at incident angle $\theta$. To perform this correction requires to know $\mu$ in absolute units. Because of the thinness of the films, the correction should be smaller than for the bulk.

\begin{figure}
    \centering
    \includegraphics[width=6.7cm]{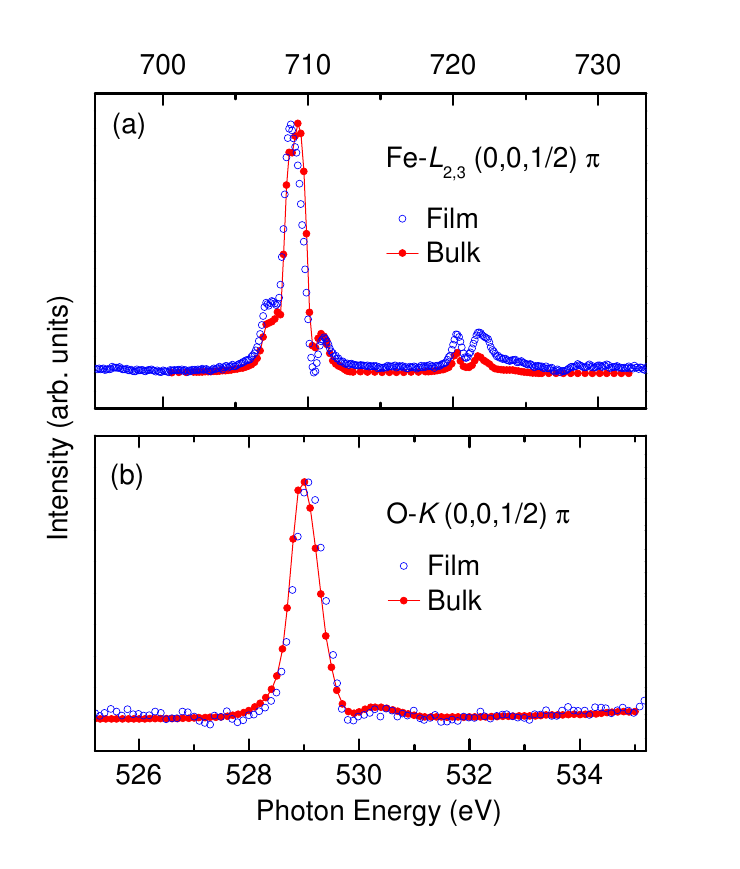}
    \caption{(color online) RSXD spectra of the
    (00$\frac{1}{2}$) peak from magnetite thin film and bulk crystal at
    (a) Fe $L_{2,3}$-edges and (b) O $K$-edge using $\pi$ polarized light.}
    \label{Fig-bulk-film}
\end{figure}

In Fig.~\ref{Fig-bulk-film} we now compare again the corrected spectra from the bulk (filled symbols) with the uncorrected film spectra as presented in Ref.~\onlinecite{schlappa:026406} (open
symbols). For the O-$K$ spectra the
agreement between bulk and film data is now close to perfect. Also
for the Fe resonance the overall agreement between the spectra is
very good, all spectral features agree. The main discrepancy here
is in the relative intensities of the $L_3$ and $L_2$ part of the
Fe-resonance spectra. The application of an absorption correction
on also the thin film data would probably resolve much of this
discrepancy. Thus, apart for this, we find very similar spectra
from the film and the bulk at both resonances. We can, hence,
safely conclude that the resonant diffraction data recorded from
the thin film sample are representative for the material magnetite
as such and that the conclusions derived from the data of
magnetite thin films in our former Letter \cite{schlappa:026406}
stand for the intrinsic properties of magnetite.

\begin{figure}
    \centering
    \includegraphics[width=6.7cm]{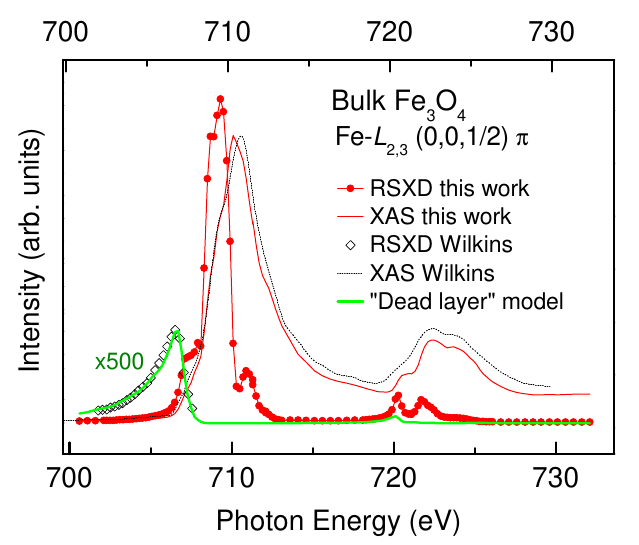}
    \caption{(color online) Fe $L_{2,3}$-edges RSXD spectra of the (00$\frac{1}{2}$)
    peak and XAS spectra taken from our magnetite bulk crystal (red)
    and from Wilkins \textit{et al.},\cite{wilkins:201102} (black). The green thick line is
    the result of a simulation multiplied by a factor 500 (see text).
    The XAS spectra are included as energy reference.}
    \label{Fig-Fe-L-with-Wilkins}
\end{figure}

We now compare our experimental data with those from Huang et
al.\cite{huang:096401}, Wilkins \textit{et al.}\cite{wilkins:201102}, and
Garc\'{i}a \textit{et al.}\cite{garcia:PRL2009} from bulk magnetite.
We first focus on the Fe $L_{2,3}$-edges data. In
Fig.~\ref{Fig-Fe-L-with-Wilkins} we present our data and those
reproduced from the paper of Wilkins \textit{et al.}\cite{wilkins:201102}.
The energy scales were aligned using the XAS data to account for
small energy-scale offsets between different soft x-ray
monochromators. The energy scale given at the top of
Fig.~\ref{Fig-Fe-L-with-Wilkins} refers to the one used by Wilkins
\textit{et al.}\cite{wilkins:201102}, and the bottom scale is the one of
our experiment. It was claimed by Wilkins \textit{et al.} that their data
were strongly distorted by absorption effects, which suppressed all scattered intensity when the absorption increases.
However, this cannot be the case because our raw data are already
very different. In fact, it is quite unlikely that absorption in
an otherwise intact sample can suppress the scattered intensity
completely. \cite{comment:absorption}

What could cause such strong distortions, however, is a layer on
top of the sample, which does absorb photons but does not
contribute to the scattered signal. This could be either a
polycrystalline layer or a crystalline layer with different
surface orientation as the bulk or a layer of different chemical
composition. That this "dead" layer scenario is likely is
demonstrated in Fig.~\ref{Fig-Fe-L-with-Wilkins}. The diamond
symbols are the experimental data from Wilkins et
al.\cite{wilkins:201102} and the green thick solid line is a
simulation in which we describe the presence of an absorbing layer
of thickness $D_a$ by multiplying our experimental RSXD spectrum
by $e^{-2d_a \mu(hv)}$ with $d_a = D_a / \sin{\theta}$.

We find the best agreement between data and simulation for $d_a
\times \mu \approx 18$ at the absorption maximum. The resulting
spectrum reproduces not only the peak position but also the shape
of the data from Wilkins \textit{et al.}\cite{wilkins:201102} very well. We
note that as compared to our experimental spectrum the green line
has been multiplied by a factor 500; more than 99 percent of the
intensity of the diffraction spectrum is taken away by the
absorbing dead layer. Concerning the Fe $L_{2,3}$ data from Garc\'{i}a
\textit{et al.}\cite{garcia:PRL2009} we note that
their (raw) data are closer to ours, although there are
appreciable quantitative differences. It therefore appears likely that the effect of an absorbing dead layer is present also in their data, although of a much thinner one than in the work of Wilkins \textit{et
al.}\cite{wilkins:201102}

\begin{figure}
    \centering
    \includegraphics[width=6.7cm]{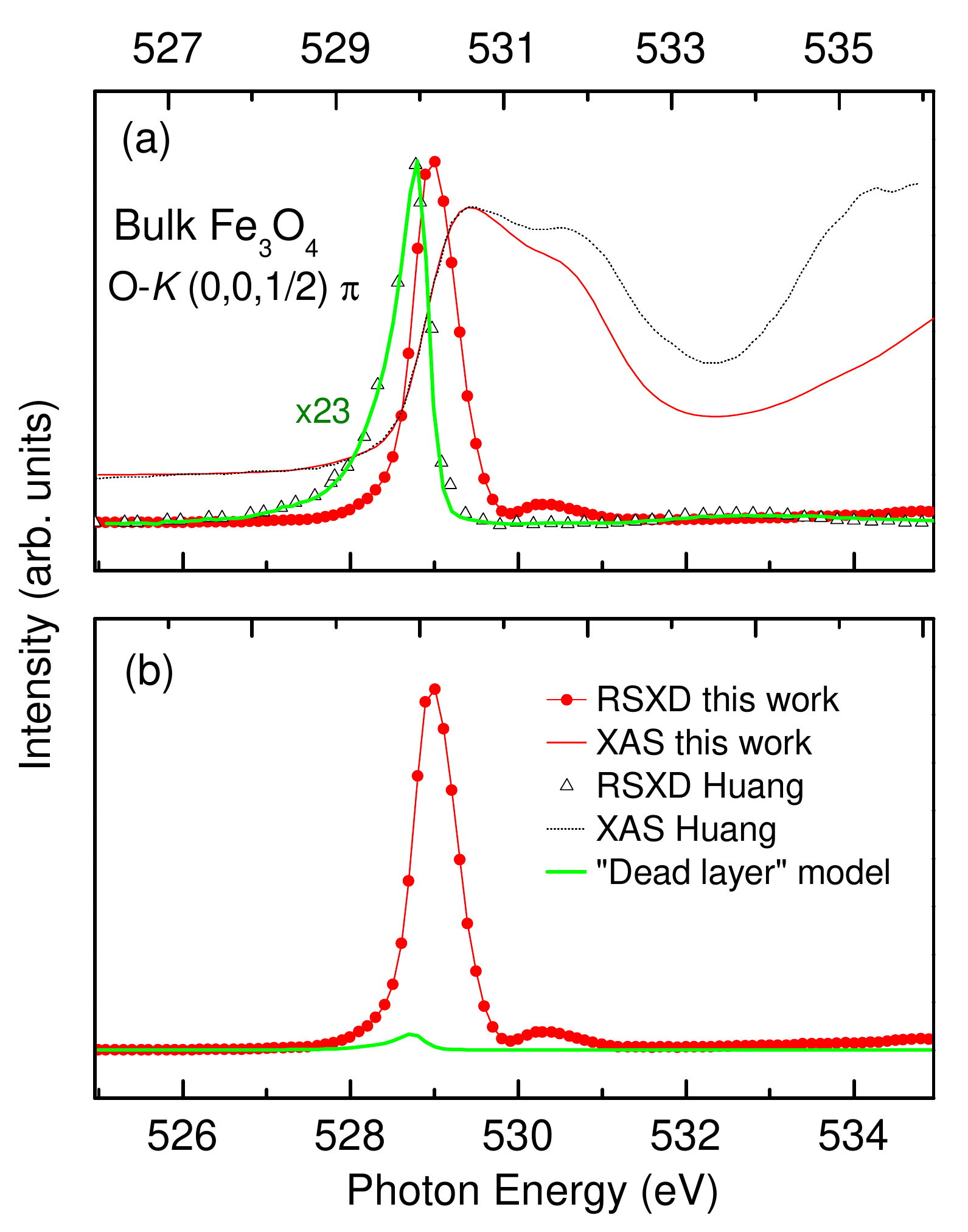}
    \caption{(color online) (a) O $K$-edge RSXD spectra of the (00$\frac{1}{2}$)
    peak and XAS spectra taken from our magnetite bulk crystal (red)
    and from Ref.~\onlinecite{huang:096401} (black). The green thick line is
    a simulated result (see text). The XAS spectra are included as energy reference.
    (b) Red line and symbols is our RSXD spectrum, and
    green thick line is the same simulated result as in (a)
    but now both on the same vertical scale, showing that more than
    95 percent of signal is taken away by the absorbing layer.}
    \label{Fig-O-K-with-Huang}
\end{figure}

We now address the oxygen resonance spectra. Experimentally both
in the papers by Wilkins \textit{et al.}\cite{wilkins:201102} and Huang et
al.\cite{huang:096401} a single resonance peak is observed, while
our data show clearly a satellite 1.3 eV higher in energy. Huang
\textit{et al.} observe additionally a broad hump at about 4 eV higher
photon energy, which is missing in our data. The 1.3-eV satellite
occurs at energies where the absorption cross section is high and
the filter effect discussed above for the Fe resonance may
suppress this feature. Since no absorption data for energy
reference are given by Wilkins \textit{et al.}\cite{wilkins:201102} we
focus on those from Huang \textit{et al.},\cite{huang:096401}. In
Fig.~\ref{Fig-O-K-with-Huang} (a) we reproduce the O-$K$ RSXD data
from Huang \textit{et al.}\cite{huang:096401}. Again, we aligned the energy
scales using the O-$K$ XAS spectra (thin/dotted lines). Since our
experiment was performed with higher energy resolution, we
broadened our experimental XAS data slightly to obtain the same
energy resolution as in the experiment of Huang et
al.\cite{huang:096401}. On such a calibrated energy scale we find
the resonance peak at a by 230 meV higher photon energy, i.e. much
closer to the absorption maximum than Huang \textit{et al.} did.

We have carried out the same simulation as in the Fe
$L_{2,3}$-edges analysis above for O $K$-resonance as shown in
Fig.~\ref{Fig-O-K-with-Huang} (a). The triangle symbols are the
experimental data from Huang \textit{et al.}\cite{huang:096401} and the
green thick solid line is the simulation result. The resulting
spectrum reproduces the main peak position and shape, and also
contains the weak and broad hump at 4 eV above the main peak, i.e.
a structure which is not present in the real spectrum and which
is artificially generated by the local minimum in the absorption
at that photon energy. In panel (b) we plot our original spectrum
(red line and symbols) and the one generated by assuming an
absorbing layer (green solid line) on the same vertical scale. One
can also see how much intensity of the diffraction spectrum is
taken away by the absorbing dead layer: more than 95 percent of
the signal is gone. The shape of the spectrum of Huang et
al.\cite{huang:096401} is completely destroyed, the 1.3-eV
satellite is suppressed, and the 4-eV feature is artificially
generated by the dead surface layer.

We note that assuming a photon mean free path ($1/\mu$) of about
1300 {\AA} (a value proposed by Huang \textit{et al.}\cite{huang:096401})
at the energy between 528 and 531 eV, and considering the
detection geometry with photon incidence and detection angle
around 45 degrees, we find a thickness of the "dead" surface
layer of about a quarter of a micrometer. A similar thickness comes out for the data of Wilkins \textit{et al.}\cite{wilkins:201102} at the Fe-$L_3$ when we estimate the photon mean free path at the Fe-$L_3$ maximum based on our XAS data.

In summary, we have recorded RSXD spectra of the (00$\frac{1}{2}$)
peak at both Fe $L_{2,3}$-edges and O $K$-edge from in-vacuo
cleaved magnetite single crystals, which essentially agree with
those from magnetite thin films published earlier. We found
that our spectra differ strongly from those published by Huang et
al.\cite{huang:096401} and Wilkins \textit{et al.}\cite{wilkins:201102},
and to lesser extent, from Garc\'{i}a \textit{et
al.}\cite{garcia:PRL2009} We were able to ascribe the distortions
in their spectra to extrinsic effects, namely the presence of a
dead surface layer. Both our bulk and thin film data can hence be used
for a reliable modeling of the charge and orbital ordering
phenomena in magnetite.

We acknowledge BESSY for excellent working conditions and Eugen
Weschke for the making his soft x-ray diffractometer available for
the experiment. We thank Lucie Hamdan and the mechanical workshop
of the II. Physikalische Institut for their skillful technical
assistance. The research in Cologne is supported by the Deutsche
Forschungsgemeinschaft through SFB 608 and by the BMBF through
contract 05KS7PK1; work at BESSY by the BMBF through contract
05ES3XBA/5.


\end{document}